\documentclass[aps,superscriptaddress,amsmath,amssymb,twocolumn,floatfix,english]{revtex4}

\usepackage{url}
\usepackage{bm}
\usepackage{graphicx}
\usepackage{amsmath}
\usepackage{amstext}
\usepackage{amssymb}
\usepackage{amsfonts}
\usepackage{amsbsy}
\usepackage{verbatim}
\usepackage{color}
\usepackage[colorlinks=true, urlcolor=blue, linkcolor=blue, citecolor=blue, pdftex]{hyperref}
\usepackage{multirow}
\usepackage{pdfpages}
\usepackage{floatrow}
\usepackage{gensymb}
\usepackage{textcomp}

\newcommand{\dagga}{{\phantom{\dagger}}}

\begin{document}

\title{Metal-insulator transitions, superconductivity, and magnetism in the two-band Hubbard model}

\author{Caterina De Franco}
\affiliation{SISSA-International School for Advanced Studies, Via Bonomea 265, I-34136 Trieste, Italy}
\author{Luca F. Tocchio}
\affiliation{Institute for Condensed Matter Physics and Complex Systems, DISAT, Politecnico di Torino, I-10129 Torino, Italy}
\author{Federico Becca}
\affiliation{Democritos National Simulation Center, Istituto Officina dei Materiali del CNR and 
SISSA-International School for Advanced Studies, Via Bonomea 265, I-34136 Trieste, Italy}

\date{\today}

\begin{abstract}
We explore the ground-state properties of the two-band Hubbard model with degenerate electronic bands, parametrized by nearest-neighbor hopping $t$, 
intra- and inter-orbital on-site Coulomb repulsions $U$ and $U^\prime$, and Hund coupling $J$, focusing on the case with $J>0$. Using Jastrow-Slater 
wave functions, we consider both states with and without magnetic/orbital order. Electron pairing can also be included in the wave function, in 
order to detect the occurrence of superconductivity for generic electron densities $n$. When no magnetic/orbital order is considered, the Mott 
transition is continuous for $n=1$ (quarter filling); instead, at $n=2$ (half filling), it is first order for small values of $J/U$, while it turns out 
to be continuous when the ratio $J/U$ is increased. A significant triplet pairing is present in a broad region around $n=2$. By contrast, singlet 
superconductivity (with $d$-wave symmetry) is detected only for small values of the Hund coupling and very close to half filling. When including 
magnetic and orbital order, the Mott insulator acquires antiferromagnetic order for $n=2$; instead, for $n=1$ the insulator has ferromagnetic and 
antiferro-orbital orders. In the latter case, a metallic phase is present for small values of $U/t$ and the metal-insulator transition becomes first 
order. In the region with $1<n<2$, we observe that ferromagnetism (with no orbital order) is particularly robust for large values of the Coulomb 
repulsion and that triplet superconductivity is strongly suppressed by the presence of antiferromagnetism. The case with $J=0$, which has an enlarged 
SU(4) symmetry due to the interplay between spin and orbital degrees of freedom, is also analyzed.
\end{abstract}

\maketitle

\section{Introduction}\label{sec:intro}

The single-band Hubbard model represents the simplest example to describe strongly-correlated systems, where the interplay between kinetic energy 
and Coulomb repulsion may give rise to a rich phase diagram, which includes insulating and conducting states, with possible superconductivity and/or 
spin/charge disproportionations~\cite{leblanc2015,zheng2017}. This model can be used to capture the low-energy properties of materials where spin 
and charge fluctuations involve predominantly one orbital (i.e., fluctuations among different orbitals are substantially quenched), as for example 
cuprate superconductors. In this regard, it is widely accepted that the single-band Hubbard model (or its strong-coupling limit, i.e., the so-called 
$t{-}J$ model) may grab the essential features of high-temperature superconductivity~\cite{dagotto1994,imada1998}. Still, there are many cases in 
which orbital fluctuations are relevant and give rise to important physical phenomena that cannot be captured within a single-band model. For example, 
the hybridization among different orbitals and the presence of the Hund coupling may produce appreciable effects at low temperatures, thus affecting 
both the normal and the superconducting phases. One of the most prominent examples is given by the iron-based superconductors, where all the $d$ 
orbitals of Iron atoms may play an important role in the conducting properties and the inclusion of multiband effects is necessary to correctly describe 
the relevant aspects of the electronic properties (e.g., the topology of the Fermi surface)~\cite{lee2008,daghofer2010,yu2011,hu2016,fernandes2017}.

Within multiband models, one key point that has been addressed in the past is to understand how the Mott metal-insulator transition (MIT) at integer 
fillings is affected by orbital degeneracy, inter-orbital Coulomb interaction, and Hund coupling. In this context, many works have been performed in 
the ``symmetric sector'', namely disregarding any possible magnetic or orbital long-range order, in order to capture the correlation effects that are 
not spoiled by weak-coupling effects. This approach is justified by the choice of describing the physical picture that can be realized when magnetic
and orbital order is suppressed by the presence of competing interactions, i.e., frustration (without including it explicitly in the model). For the 
single-band Hubbard model, this way of proceeding has been widely used within the Gutzwiller approximation~\cite{brinkman1970,vollhardt1984}, dynamical 
mean-field theory (DMFT)~\cite{georges1996}, slave-boson approaches~\cite{kotliar1986}, and variational Monte Carlo methods~\cite{capello2006}. For 
the $M$-band Hubbard model, in the absence of the Hund coupling $J$, it has been observed that the value of the Coulomb interaction $U_{\rm MIT}$, for 
which the MIT occurs at commensurate filling $n$, reaches its maximum at half filling, i.e., for $n=M$. This result has been obtained by using the 
Gutzwiller approximation~\cite{lu1994}, DMFT~\cite{rozenberg1997,ono2003}, and quantum Monte Carlo techniques~\cite{gunnarsson1999}. The presence of 
a finite $J$ term reduces the value of $U_{\rm MIT}$ at half filling~\cite{ono2003,han1998}. Then, recent studies~\cite{demedici2011a,demedici2011b} 
highlighted the opposite trend for all the other (integer) fillings, where the presence of a finite $J$ increases $U_{\rm MIT}$ (in this case, the 
existence of a correlated metal with tiny quasiparticle weight has been also emphasized~\cite{demedici2011a,demedici2011b,fanfarillo2015,nomura2015}). 
One important issue that has been addressed in multiband Hubbard models is the nature of the MIT. Indeed, while in the single-band model different 
numerical methods~\cite{brinkman1970,georges1996,capello2005} established that the MIT is continuous at zero temperature, former studies of multiband 
models, based on the Gutzwiller approximation, suggested that the transition, at half filling, becomes first order whenever $J>0$, while it remains 
continuous only at $J=0$~\cite{buenemann1997,buenemann1998}. Similar results have been obtained more recently by means of the DMFT 
method~\cite{ono2003,facio2017}. 

The analysis of the role of band degeneracy and Hund coupling in the development of superconductivity in multiband Hubbard models represents another 
topic of great interest, particularly relevant for iron-based superconductors. However, treating nonlocal pairing beyond perturbative approximations 
is particularly difficult. A recent DMFT study on a three-band Hubbard model highlighted the emergence of on-site (i.e., local) triplet superconductivity 
at finite doping for $J>0$~\cite{hoshino2015}, in agreement with previous results obtained in the large $J/U$ limit, within an Hartree-Fock-Bogoliubov 
approach~\cite{zegrodnik2012} and the Gutzwiller approximation~\cite{zegrodnik2013}. Here, spin-triplet superconductivity is related to the emergence 
of local magnetic moments, which originate from the Hund coupling and are enhanced by an Ising anisotropy that suppresses fluctuations among different 
spin configurations. The presence of non-local pairing (i.e., with $d$-wave symmetry) is much more difficult to assess within DMFT, since it would
require a cluster extension, which is computationally heavy for multiorbital systems. 

In addition to superconductivity, long-range magnetic order may be stabilized in a relatively large region of the phase diagram for $J>0$. Within the
two-band model, various calculations highlighted the existence of itinerant ferromagnetism for $1<n<2$, which can be stabilized by the double-exchange 
mechanism for $J>0$~\cite{held1998,momoi1998,kubo2009,peters2010}. In addition, recent DMFT calculations on the three-band model~\cite{hoshino2015} 
suggested the possibility to have antiferromagnetism close to half filling and ferromagnetism in a wide doping region at large values of the Coulomb 
repulsion. In the $J=0$ limit, the situation is delicate; in fact, the model with degenerate bands possesses an enlarged SU$(2M)$ symmetry, which is 
generated by spin and orbital degrees of freedom. Models with SU$(N)$ symmetry have been investigated within the strong-coupling limit, i.e., within 
the Heisenberg model~\cite{hermele2011}. In the square lattice for $N=4$ (corresponding to two electronic bands), a variety of numerical calculations 
suggested the presence of a spontaneous symmetry breaking in the ground state, for both one and two particles per site~\cite{corboz2011,kim2017}. 
Within the Hubbard model for $N>2$, quantum fluctuations could be sufficiently strong to destroy magnetic/orbital order at small values of the Coulomb 
interaction even at half filling in the presence of a perfect nesting of the underlying Fermi surface (instead, for $N=2$, the ground state has 
long-range magnetic order for any value of $U$ at half filling). In the weak- and intermediate-coupling limit, the cases with $N=4$ and $6$ have been 
considered in a generalized Hubbard-Heisenberg model at half filling, suggesting that for $U=0$ and small values of the antiferromagnetic coupling 
a $d$-density wave state is stabilized~\cite{assaad2005}. 

In this paper, we consider the two-band Hubbard model with degenerate bands on a square lattice, as the simplest case to investigate the role of the 
inter-orbital Coulomb repulsion and Hund coupling, while keeping the band structure as simple as possible, i.e., with only nearest-neighbor hopping.
The same band structure has been widely considered in the past and represents the first step to generalize the single-band Hubbard model toward 
the multiband case~\cite{georges2013}. We analyze the model by means of the variational Monte Carlo method. This approach, which works 
directly in two spatial dimensions, allows us to present a point of view that is complementary with respect to previous DMFT studies, which apply to 
infinite dimensions. First of all, we locate the MIT at commensurate fillings when no magnetic/orbital order is considered within the variational 
{\it Ansatz}. For the generic case with $J>0$, the Mott transition appears to be continuous for $n=1$; instead, for $n=2$, it is first order for small 
values of $J/U$ and turns out to be continuous when the Hund coupling is increased. At half filling, the Mott transition is also accompanied by the 
stabilization of a sizable on-site triplet pairing, which survives in a wide region of doping around $n=2$. A small singlet pairing with $d$-wave 
symmetry is also observed in a narrow region close to $n=2$ for sufficiently small values of the Hund coupling. A finite singlet pairing can be 
stabilized also for $J=0$, thus breaking the SU(4) symmetry in the variational wave function; in this case the MIT is first order; by contrast, when 
a fully-symmetric {\it Ansatz} is considered, the Mott transition becomes continuous. At quarter filling and close to it, neither triplet nor singlet 
pairing can be stabilized, indicating that superconductivity is not present around $n=1$ in the two-band Hubbard model with degenerate bands.

Symmetry-breaking states can be studied by including magnetic/orbital order within the variational {\it Ansatz}. At half filling, antiferromagnetic 
order is stabilized for small and intermediate values of the Coulomb interaction $U$ even for $J=0$, suggesting that the SU(4) symmetry can be broken 
at small values of $U/t$. At quarter filling, the metallic phase is stable for small values of $U/t$, while the Mott insulator acquires both 
ferromagnetic and antiferro-orbital orders, in agreement with previous calculations~\cite{held1998,momoi1998,kubo2009,peters2010}. For $1<n<2$,
a wide region of ferromagnetism (without orbital order) is found for large values of the Coulomb repulsion. For small values of the Hund coupling, 
phase separation for $n>1$ may appear. By contrast, the region of stability of antiferromagnetism is limited to dopings close to $n=2$. In the 
presence of magnetic order, triplet superconductivity is strongly suppressed close to half filling, coexisting with antiferromagnetic order.

The rest of the paper is organized as follows: In Sec.~\ref{sec:methods}, we introduce the two-band Hubbard model and the variational wave functions 
that are used within the Monte Carlo method. In Sec.~\ref{sec:results}, we describe our results on the metal-insulator transitions, superconductivity, 
and magnetic/orbital order. Finally, in Sec.~\ref{sec:conclusions}, we draw our conclusions. 

\section{Model and method}\label{sec:methods}

We consider the two-band Hubbard model defined by:
\begin{equation}\label{eq:hamtot}
{\cal H} = {\cal H}_{\textrm{kin}} + {\cal H}_{\textrm{int}},
\end{equation}
where the kinetic term ${\cal H}_{\textrm{kin}}$ describes hopping processes of electrons within two degenerate  orbitals:
\begin{equation}\label{eq:hamkin}
{\cal H}_{\textrm{kin}} = -t \sum_{\langle i,j\rangle,\alpha,\sigma} 
c^\dagger_{i,\alpha,\sigma} c^\dagga_{j,\alpha,\sigma} + \rm{H.c.};
\end{equation}
here $c^\dagger_{i,\alpha,\sigma}$ ($c^\dagga_{i,\alpha,\sigma}$) creates (destroys) an electron with spin $\sigma$ on site $i$ 
and orbital $\alpha=1,2$ and $t$ is the nearest-neighbor hopping amplitude on the square lattice. The interaction term includes 
four different contributions:
\begin{eqnarray}
{\cal H}_{\textrm{int}} &=& U \sum_{i,\alpha} n_{i,\alpha,\uparrow}n_{i,\alpha,\downarrow}
                        + U^\prime \sum_{i,\sigma,\sigma^\prime} n_{i,1,\sigma}n_{i,2,\sigma^\prime} \nonumber \\
                        &-& J \sum_{i,\sigma,\sigma^\prime} c^\dagger_{i,1,\sigma} c^\dagga_{i,1,\sigma^\prime}
                                                            c^\dagger_{i,2,\sigma^\prime} c^\dagga_{i,2,\sigma} \nonumber \\
                        &-& J^\prime \sum_{i} (c^\dagger_{i,1,\uparrow} c^\dagger_{i,1,\downarrow}
                                               c^\dagga_{i,2,\uparrow} c^\dagga_{i,2,\downarrow} + \rm{H.c.}),
\label{eq:hamint}
\end{eqnarray}
where $n_{i,\alpha,\sigma}=c^\dagger_{i,\alpha,\sigma} c^\dagga_{i,\alpha,\sigma}$ is the electronic density per spin on site $i$ and orbital 
$\alpha$. These four terms represent the intra-orbital (inter-orbital) Coulomb interaction $U$ ($U^\prime$) and the spin-flip (pair-hopping) 
Hund term $J$ ($J^\prime$). In the following, we set $U^\prime=U-2J$ and $J^\prime=J$~\cite{georges2013}.

Our numerical results are obtained by means of the variational Monte Carlo method, which is based on the definition of suitable 
wave functions to approximate the ground-state properties beyond perturbative approaches. In particular, we consider the so-called 
Jastrow-Slater wave functions that extend the original formulation proposed by Gutzwiller to include correlations effects on top
of uncorrelated states~\cite{gutzwiller1963,yokoyama1987}. Our variational states are described by:
\begin{equation}\label{eq:WF}
|\Psi\rangle={\cal J}|\Phi_0\rangle;
\end{equation}
here, ${\cal J}$ is the density-density Jastrow factor, which is defined by:
\begin{equation}\label{eq:jastrow}
{\cal J} = \exp \left ( -\frac{1}{2} \sum_{i,j,\alpha,\beta} 
v^{\alpha,\beta}_{i,j} n_{i,\alpha} n_{j,\beta} \right ),
\end{equation}
where $n_{i,\alpha}= \sum_{\sigma} n_{i,\alpha,\sigma}$ is the electron density on site $i$ and orbital $\alpha$; 
$v^{\alpha,\beta}_{i,j} = v^{\beta,\alpha}_{i,j}$ (that include also the local Gutzwiller term for $\alpha=\beta$ and $i=j$) 
are pseudopotentials that are optimized for every independent distance $|{\bf R}_i-{\bf R}_j|$. In the following, we will 
consider $v^{1,1}_{i,j}=v^{2,2}_{i,j} \equiv v^{\rm intra}_{i,j}$ and $v^{1,2}_{i,j}=v^{2,1}_{i,j} \equiv v^{\rm inter}_{i,j}$. 
Moreover, the Fourier transform of the intra- and inter-orbital Jastrow terms will be denoted by $v^{\rm intra}(\bm{q})$ and 
$v^{\rm inter}(\bm{q})$, respectively. The Jastrow factor has been shown to be crucial in describing a Mott insulating state 
within the single-band Hubbard model~\cite{capello2005}. As far as the two-band Hubbard model is concerned, the role of the 
Jastrow factor has been already highlighted in a variational Monte Carlo study of the orbital-selective Mott 
insulator~\cite{tocchio2016a} and of the square lattice bilayer Hubbard model~\cite{rueger2014}. Then, $|\Phi_0\rangle$ is 
an uncorrelated state that is constructed from an auxiliary (quadratic) Hamiltonian:
\begin{equation}\label{eq:hamaux}
{\cal H}_{\textrm{aux}} = {\cal H}_{\textrm{kin}} + {\cal H}_{\textrm{sc}} + {\cal H}_{\textrm{mag}} + {\cal H}_{\textrm{orb}},
\end{equation} 
where ${\cal H}_{\textrm{kin}}$ is the kinetic term defined in Eq.~(\ref{eq:hamkin}), ${\cal H}_{\textrm{sc}}$ includes electron
pairing and chemical potential:
\begin{eqnarray}
{\cal H}_{\textrm{sc}} &=& \sum_{\langle i,j \rangle,\alpha} \Delta_{i,j}
                           \left (c^\dagger_{i,\alpha,\uparrow} c^\dagger_{j,\alpha,\downarrow}
                                + c^\dagger_{j,\alpha,\uparrow} c^\dagger_{i,\alpha,\downarrow} \right ) + \rm{H.c.} \nonumber \\
                       &+& \Delta_{\perp} \sum_{i} 
                           \left( c^\dagger_{i,1,\uparrow} c^\dagger_{i,2,\downarrow}
                                - c^\dagger_{i,2,\uparrow} c^\dagger_{i,1,\downarrow} \right) + \rm{H.c.} \nonumber \\
                       &+& \mu \sum_{i,\alpha,\sigma} c^\dagger_{i,\alpha,\sigma} c^\dagga_{i,\alpha,\sigma};
\label{eq:hamBCS}
\end{eqnarray}
${\cal H}_{\textrm{mag}}$ and ${\cal H}_{\textrm{orb}}$ incorporate magnetic and orbital orders:
\begin{eqnarray}
\label{eq:hamMAG}
{\cal H}_{\textrm{mag}} &=& \Delta_{\rm AFM} \sum_{i,\alpha} (-1)^{{\bf R}_i} \left( 
                            c^\dagger_{i,\alpha,\uparrow} c^{\dagga}_{i,\alpha,\uparrow} 
                          - c^\dagger_{i,\alpha,\downarrow} c^\dagga_{i,\alpha,\downarrow} \right) \nonumber \\
                        &+& h_{\rm FM} \sum_{i,\alpha} \left(
                            c^\dagger_{i,\alpha,\uparrow} c^{\dagga}_{i,\alpha,\uparrow}
                          - c^\dagger_{i,\alpha,\downarrow} c^\dagga_{i,\alpha,\downarrow} \right), \\
{\cal H}_{\textrm{orb}} &=& \Delta_{\rm AFO} \sum_{i,\sigma} (-1)^{{\bf R}_i} \left( 
                            c^\dagger_{i,1,\sigma} c^{\dagga}_{i,1,\sigma} 
                          - c^\dagger_{i,2,\sigma} c^\dagga_{i,2,\sigma} \right) \nonumber \\
                        &+& h_{\rm FO} \sum_{i,\sigma} \left(
                            c^\dagger_{i,1,\sigma} c^{\dagga}_{i,1,\sigma}
                          - c^\dagger_{i,2,\sigma} c^\dagga_{i,2,\sigma} \right).
\label{eq:hamORB}
\end{eqnarray}

%%%%%%%%%%%%%%%%%%%%%%%%%%%%%%%%%%%%%%%%%%%%%
\begin{figure}
\includegraphics[width=1.0\columnwidth]{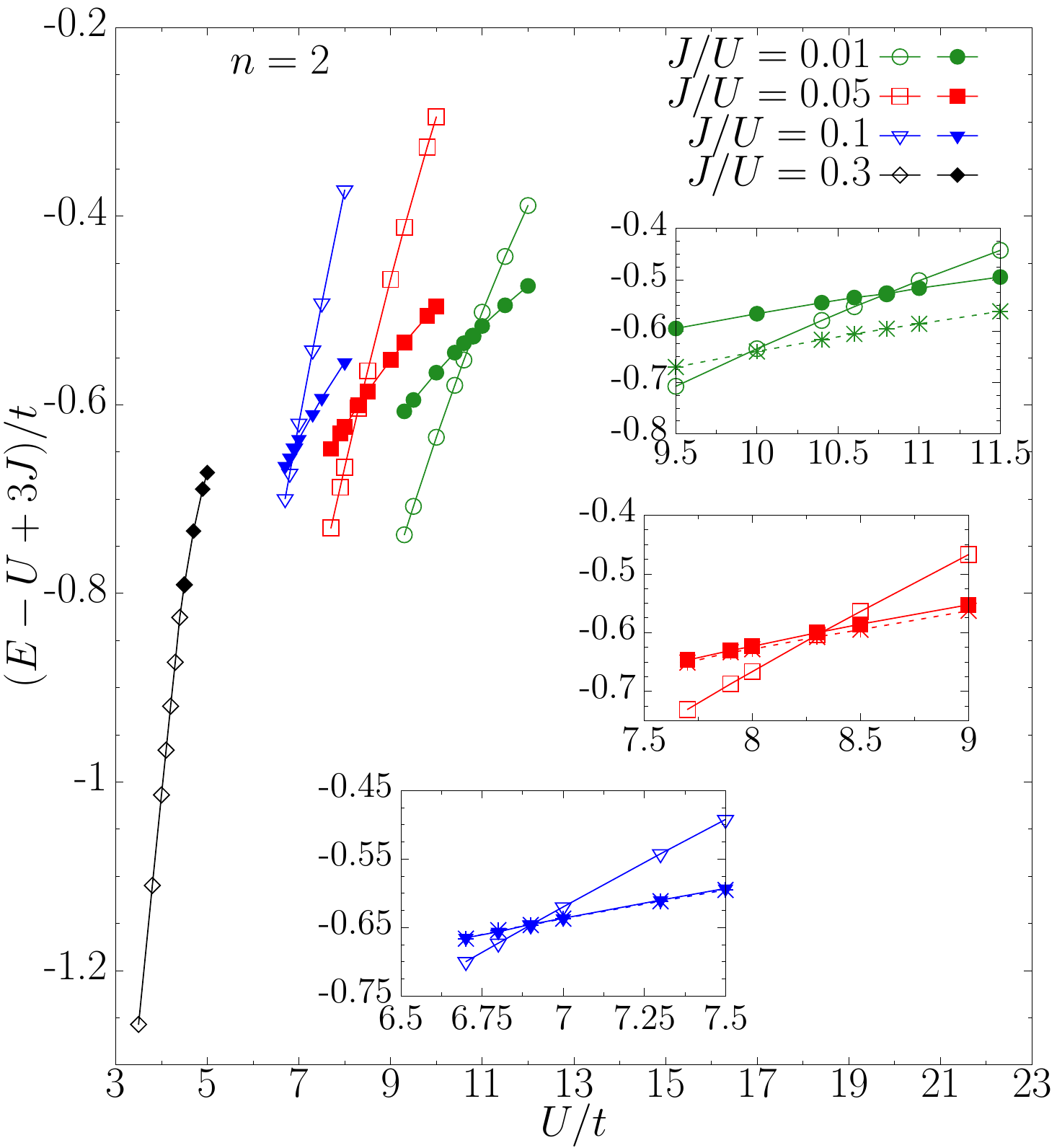}
\caption{\label{fig:mitn2}
Main panel: Energies (per site) of the metallic (empty symbols) and insulating (full symbols) states as a function of $U/t$ for 
$n=2$ and different values of the Hund coupling $J$; for clarity a constant shift of $U-3J$, which is the ground-state energy 
(per site) in the $U/t \to \infty$ limit, has been considered. For $J/U=0.01$, $0.05$, and $0.1$ the transition is first order, 
while for $J/U=0.3$ it is continuous. No magnetic or orbital orders are considered within the variational wave functions. Insets: 
zooms around the metal-insulator transitions. The stars denote the energies of the insulating state when $\Delta_d$ is allowed 
in the variational state.} 
\end{figure}
%%%%%%%%%%%%%%%%%%%%%%%%%%%%%%%%%%%%%%%%%%%%%

In Eq.~(\ref{eq:hamBCS}), $\Delta_{\perp}$ describes (on-site inter-orbital) triplet pairing, $\Delta_{i,j}$ (nearest-neighbor 
intra-orbital) singlet pairing with $d_{x^2-y^2}$ symmetry, namely $\Delta_k=2 \Delta_d [\cos(k_x) - \cos(k_y)]$ is its Fourier 
transform. In Eqs.~(\ref{eq:hamMAG}) and~(\ref{eq:hamORB}), $\Delta_{\rm AFO}$, $h_{\rm FM}$, $\Delta_{\rm FMO}$, and $h_{\rm FO}$ 
represent staggered and uniform parameters for magnetic and orbital orders. All these terms are further variational parameters 
that may be optimized in order to minimize the variational energy.

In the generic case with a finite Hund coupling $J$, wave functions with no magnetic/orbital order can be constructed by considering
only ${\cal H}_{\textrm{kin}}$ and ${\cal H}_{\textrm{sc}}$. Notice that the latter one breaks the spin SU(2) symmetry whenever a 
triplet pairing is considered, without necessarily leading to a magnetic order. As far as the Jastrow factor is concerned, for the 
generic case with a finite Hund coupling, different intra- and inter-orbital pseudopotentials are allowed in Eq.~(\ref{eq:jastrow}), 
i.e., $v^{\rm intra}_{i,j} \ne v^{\rm inter}_{i,j}$. For $J=0$, a fully-symmetric wave function requires no pairing terms in 
Eq.~(\ref{eq:hamaux}), i.e., only ${\cal H}_{\textrm{kin}}$ can be retained in the auxiliary Hamiltonian, and a Jastrow factor that 
only involves the total electron density on each site, i.e., $v^{\rm intra}_{i,j}=v^{\rm inter}_{i,j}$. Finally, states with magnetic 
and/or orbital order (with either staggered or uniform patterns) are easily obtained by also including ${\cal H}_{\textrm{mag}}$ 
and/or ${\cal H}_{\textrm{orb}}$.

In order to assess the metallic or insulating nature of the ground state, we calculate the density-density structure factor $N(q)$, 
defined as:
\begin{equation}\label{eq:nqnq}
N(\bm{q}) = \frac{1}{L} \sum_{i,j} \sum_{\alpha,\beta} \langle n_{i,\alpha} n_{j,\beta} \rangle e^{i\bm{q}\cdot(\bm{R}_i-\bm{R}_j)},
\end{equation}
where $\langle \dots \rangle$ indicates the expectation value over the variational wave function of Eq.~(\ref{eq:WF}). Indeed, a 
metallic phase has $N(\bm{q}) \propto |\bm{q}|$ for $|\bm{q}| \to 0$, corresponding to the existence of gapless excitations, while an 
insulator is expected to have $N(\bm{q}) \propto |\bm{q}|^2$~\cite{feynman1954,tocchio2011}. Within our definition of the 
variational wave function, the metallic or insulating character can be also detected by looking at the small-$q$ behavior of the Jastrow 
factor, as shown within the single-band Hubbard model~\cite{capello2005}. In the two-band Hubbard model, the metallic phase is described 
by $v^{\rm intra}(\bm{q}) \propto 1/|\bm{q}|$ [and $v^{\rm inter}(\bm{q}) \propto 1/|\bm{q}|$], while the Mott insulating phase has 
instead $v^{\rm intra}(\bm{q}) \propto 1/|\bm{q}|^2$ (and $v^{\rm inter}(\bm{q}) \propto 1/|\bm{q}|^2$).

%%%%%%%%%%%%%%%%%%%%%%%%%%%%%%%%%%%%%%%%%%%%%
\begin{figure}
\includegraphics[width=1.0\columnwidth]{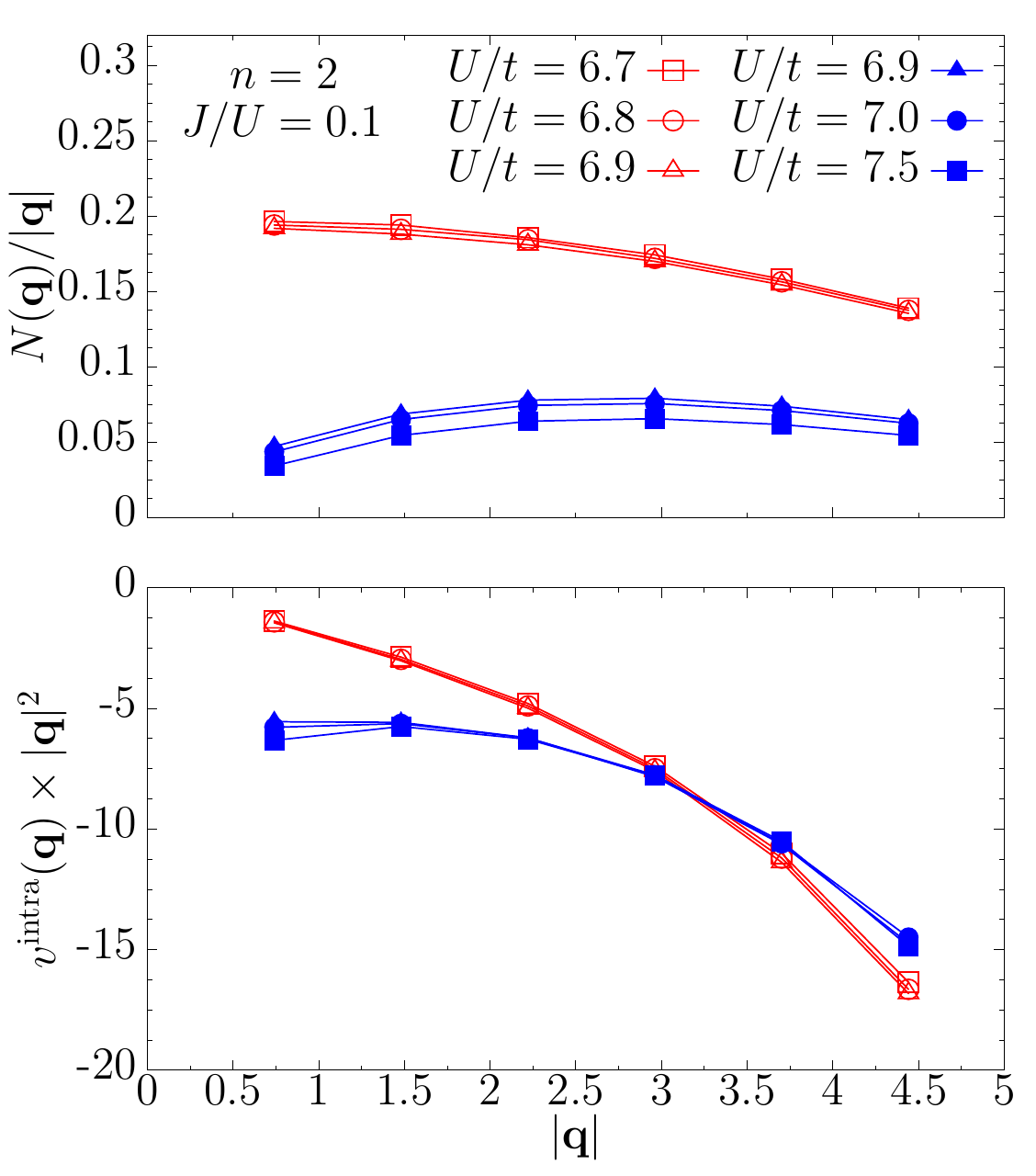}
\caption{\label{fig:Nqn2}
Results for $n=2$ and $J/U=0.1$. Upper panel: Density-density structure factor $N(\bm{q})$ of Eq.~(\ref{eq:nqnq}) (divided by $|{\bf q}|$) 
for various values of $U/t$. Lower panel: The Fourier transform of the intra-orbital Jastrow factor $v^{\rm intra}({\bf q})$ (multiplied 
by $|{\bf q}|^2$) for the same set of parameters as in the upper panel. The results for metallic (insulating) wave functions are denoted 
by empty (full) symbols. No magnetic or orbital orders are considered within the variational wave functions.}
\end{figure}

\begin{figure}
\includegraphics[width=1.0\columnwidth]{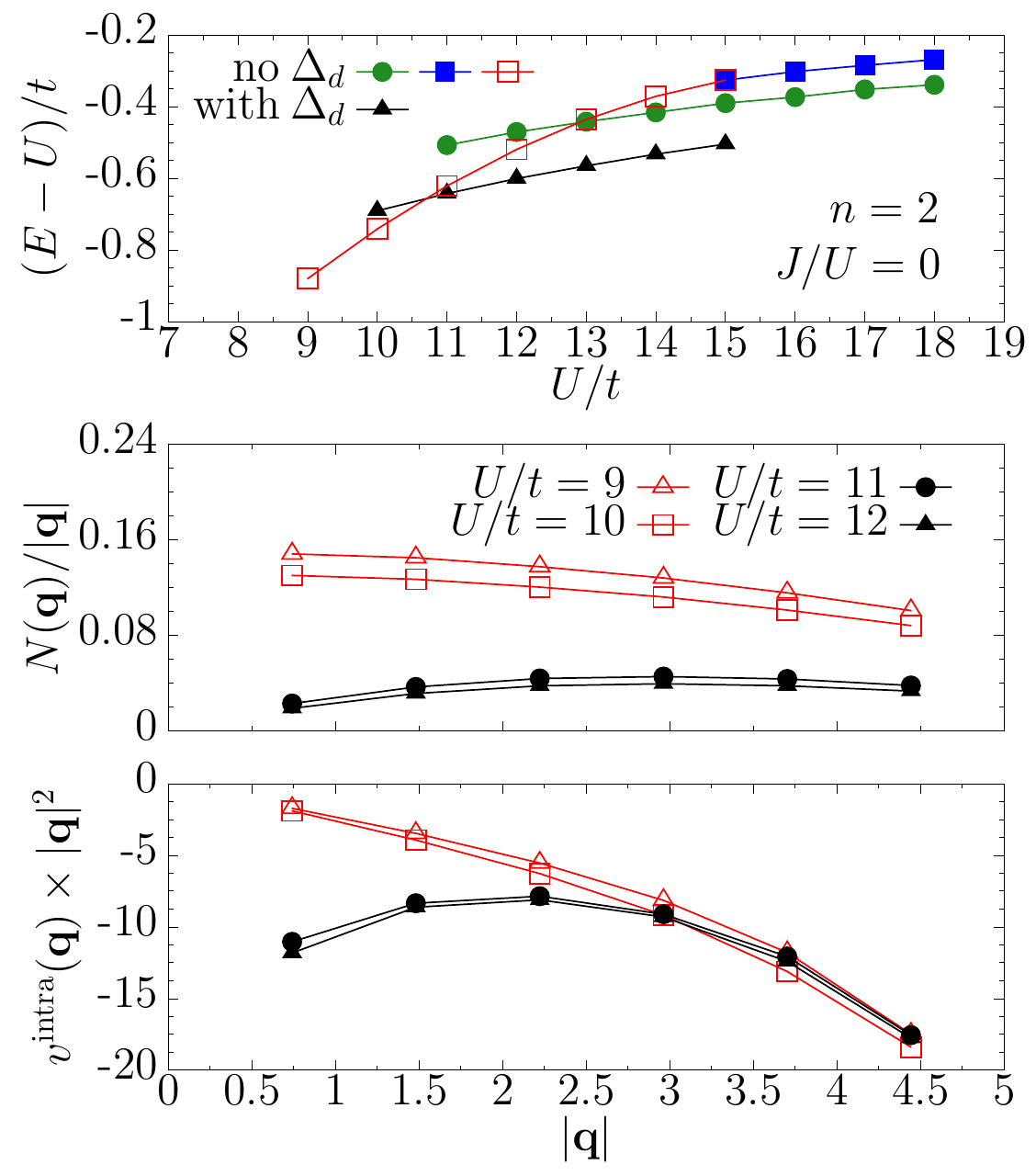}
\caption{\label{fig:mitn2J0}
Results for $n=2$ and $J=0$. Upper panel: Variational energies (per site) for the metallic and insulating states; for clarity a constant 
shift of $U$ has been considered. Metallic and insulating wave functions that do not break the SU(4) symmetry are denoted by empty and full 
squares, respectively. The insulating state with no pairing but different intra- and inter-orbital Jastrow parameters is denoted by full 
circles; finally, the insulating state with a finite $\Delta_d$ is also reported (full triangles). Middle panel: The density-density structure 
factor $N(\bm{q})$ of Eq.~(\ref{eq:nqnq}) (divided by $|{\bf q}|$) at various values of $U/t$, for the best variational state. Lower panel: 
The Fourier transform of the intra-orbital Jastrow factor (multiplied by $|{\bf q}|^2$) for the same set of parameters as in the middle panel. 
No magnetic or orbital orders are considered within the variational wave functions.}
\end{figure}
%%%%%%%%%%%%%%%%%%%%%%%%%%%%%%%%%%%%%%%%%%%%%

\section{Results}\label{sec:results}

In this section, we show our main results for the metal-insulator transitions at half filling ($n=2$) and quarter filling ($n=1$), 
including the case where magnetic and orbital orders are prevented, and for superconductivity for densities between $n=1$ and $2$.
Most of the calculations are performed on the $12 \times 12$ cluster with periodic (antiperiodic) boundary conditions along the $x$ ($y$) 
direction, in order to have a non-degenerate ground state for $U=J=0$.

\subsection{The metal-insulator transitions without magnetic/orbital orders}

Let us start to study the MIT at commensurate electron densities, $n=2$ and $n=1$, when no magnetic/orbital orders are allowed 
within the variational wave function. The results for $n=2$ and $J>0$ are reported in Fig.~\ref{fig:mitn2}. For $J/U=0.01$, $0.05$, 
and $0.1$, the Mott transition is first order, since two different wave functions, whose energies cross at $U=U_{\rm MIT}$, can be 
stabilized (in the vicinity of the MIT). While for small values of the Coulomb interaction, the best variational state is metallic 
with $N(\bm{q}) \propto |\bm{q}|$ in the limit of $|\bm{q}| \to 0$, for large $U/t$, the lowest-energy state is insulating with 
$N(\bm{q}) \propto |\bm{q}|^2$, see Fig.~\ref{fig:Nqn2}. This modification in the density-density correlations is triggered by the 
Jastrow factor, e.g., $v^{\rm intra}(\bm{q}) \propto 1/|\bm{q}|$ in the metal, while $v^{\rm intra}(\bm{q}) \propto 1/|\bm{q}|^2$
in the insulator, see Fig.~\ref{fig:Nqn2}. 

We mention that the region where metastable solutions can be stabilized shrinks as $J$ increases, thus suggesting that the transition 
may become second order for a large enough value of the Hund coupling, see also Refs.~\cite{lechermann2007,demedici2017a,facio2017}. 
Indeed, for $J/U=0.3$, the MIT appears to be continuous, with no metastable solutions that can be obtained, see Fig.~\ref{fig:mitn2}. 
Still, the small-$q$ behavior of the Jastrow factor is different for $U<U_{\rm MIT}$ and $U>U_{\rm MIT}$, as in the single-band 
Hubbard model, where the Mott transition is continuous~\cite{capello2006}.  Furthermore, our variational approach reproduces the 
well-known fact that $U_{\rm MIT}$ decreases with increasing $J$, since the Mott insulator with localized moments may take advantage 
of the Hund coupling~\cite{ono2003,han1998}.

Within the metallic regime, there is no appreciable gain when including superconducting pairing (either singlet or triplet) in the 
auxiliary Hamiltonian of Eq.~(\ref{eq:hamaux}); a similar result has been obtained in the paramagnetic solution of the single-band 
Hubbard model, where the metallic phase at half filling has vanishingly small pairing correlations~\cite{tocchio2012,dayal2012}. 
In addition, the intra- and inter-orbital Jastrow pseudopotentials are approximately equal for every distance, indicating that the 
correlation between two electrons on the same orbital is similar to the one between two electrons on different orbitals. By contrast, 
within the insulating phase, the intra-orbital Jastrow factor is larger than the inter-orbital one, implying that configurations with 
two electrons on the same orbital are penalized with respect to the ones with two electrons on different orbitals, as expected in the 
presence of a finite value of $J$. Only for small values of $J/U$, a (nearest-neighbor intra-orbital) singlet pairing with $d_{x^2-y^2}$ 
symmetry can be stabilized (see Fig~\ref{fig:mitn2}), similarly to what occurs in the single-band Hubbard model at half 
filling~\cite{capello2006,tocchio2012}. Most importantly, a strong (on-site inter-orbital) triplet pairing $\Delta_{\perp}$ is stabilized 
by the presence of a finite Hund coupling, giving a sizable gain in the variational energy with respect to the case with no pairing (see 
also Sec.~\ref{sec:superc}). Nonetheless, we must emphasize that the Jastrow factor with $v^{\rm intra}(\bm{q}) \propto 1/|\bm{q}|^2$ 
and $v^{\rm inter}(\bm{q}) \propto 1/|\bm{q}|^2$ is able to destroy the superconducting long-range order that is present in the 
uncorrelated wave function $|\Phi_0\rangle$~\cite{capello2006}. Therefore, the presence of electronic pairing in $|\Phi_0\rangle$ leads 
to the existence of ``preformed pairs'' without phase coherence in the full correlated wave function $|\Psi\rangle$ of Eq.~(\ref{eq:WF}), 
as in the single-band Hubbard model. The relevant difference with respect to the latter case is that here ``preformed pairs'' do not form 
singlets with $d_{x^2-y^2}$ symmetry, but triplets with $s$ (on-site) symmetry.

%%%%%%%%%%%%%%%%%%%%%%%%%%%%%%%%%%%%%%%%%%%%%
\begin{figure}
\includegraphics[width=1.0\columnwidth]{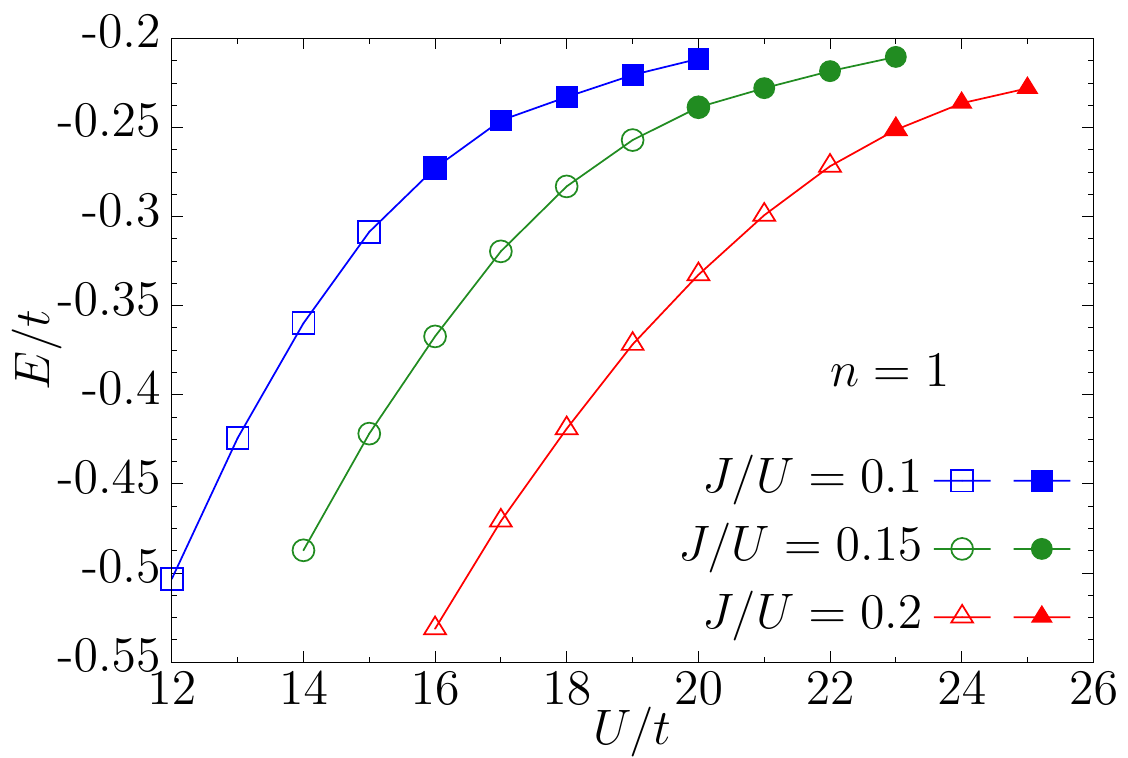}
\caption{\label{fig:mitn1}
Energies (per site) of the metallic (empty symbols) and insulating (full symbols) states as a function of $U/t$ for $n=1$ and 
different values of the Hund coupling $J$. No magnetic or orbital orders are considered within the variational wave functions.}
\end{figure}

\begin{figure}
\includegraphics[width=1.0\columnwidth]{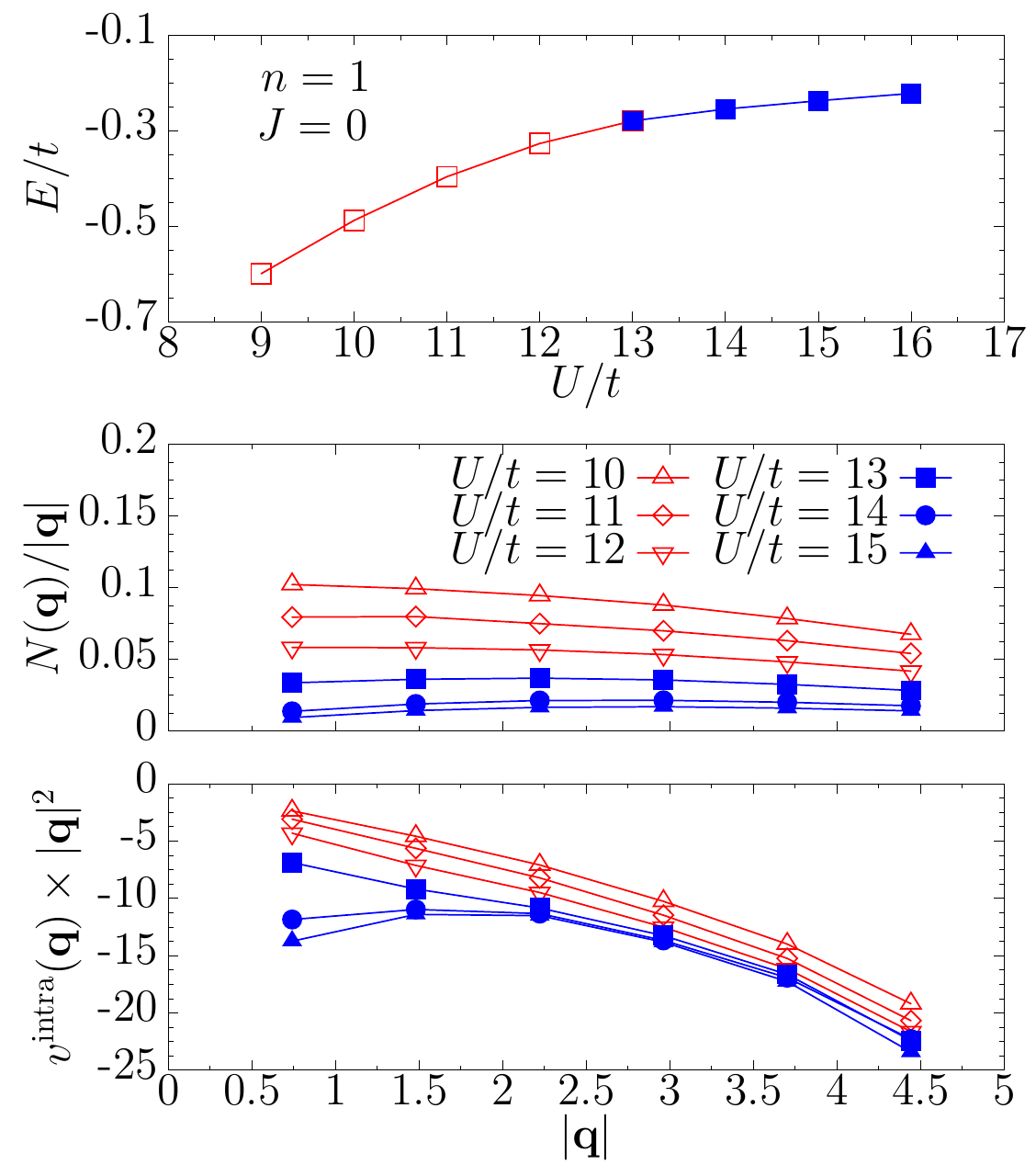}
\caption{\label{fig:Nqn1J0}
Results for $n=1$ and $J=0$. Upper panel: Variational energies for the metallic (empty symbols) and insulating (full symbols) states.
Middle panel: Density-density structure factor $N(\bm{q})$ of Eq.~(\ref{eq:nqnq}) (divided by $|{\bf q}|$) for various values of $U/t$. 
The results for metallic and insulating wave functions are denoted by empty and full symbols, respectively. Lower panel: The Fourier 
transform of the intra-orbital Jastrow factor (multiplied by $|{\bf q}|^2$) for the same set of parameters as in the middle panel. No 
magnetic or orbital orders are considered within the variational wave functions.}
\end{figure}
%%%%%%%%%%%%%%%%%%%%%%%%%%%%%%%%%%%%%%%%%%%%%

We now briefly discuss the case with $J=0$ at $n=2$. Here, whenever the variational wave function is taken to have a full SU(4) symmetry 
(i.e., by only considering the kinetic term in the auxiliary Hamiltonian and imposing $v^{\rm intra}_{i,j} = v^{\rm inter}_{i,j}$), the
transition appears to be continuous (at $U_{\rm MIT}/t=15 \pm 1$), with no metastable solutions in the energy optimization, see
Fig.~\ref{fig:mitn2J0}. However, by allowing different intra- and inter-orbital Jastrow factors in the variational optimization, another 
insulating solution exists, which is energetically favorable for $U/t \gtrsim 13$, see Fig.~\ref{fig:mitn2J0}. Then, this insulating state 
can be further improved by considering the electron (singlet) pairing (with $d_{x^2-y^2}$ symmetry) in the auxiliary Hamiltonian, further
lowering the transition to $U_{\rm MIT}/t=11 \pm 0.5$. As before, the Jastrow factor prevents the existence of off-diagonal superconducting
order.

Let us now investigate the case with $n=1$, for which the results are shown in Fig.~\ref{fig:mitn1}. In contrast to the half-filled case, 
here the Mott transition is always continuous and is marked by a progressive change in the small-$q$ behavior of the Jastrow factor, see
Fig.~\ref{fig:Nqn1J0}. Remarkably, no gain in the variational energy is detected by allowing (on-site inter-orbital) triplet or 
(nearest-neighbor intra- or inter-orbital) singlet pairings, both in the metallic and the insulating phases. In addition, the intra- and
inter-orbital Jastrow pseudopotentials are very similar, implying that the variational wave function remains fully symmetric not only for 
$J=0$ but also for $J>0$. In particular, for the former case, we find that $U_{\rm MIT}/t= 13 \pm 1$. This result indicates that, within 
SU(4) symmetric solutions, the maximum value of $U_{\rm MIT}$ is obtained at half filling, in agreement with previous DMFT and Gutzwiller 
approximation calculations~\cite{lu1994,rozenberg1997,ono2003}. Instead, when we allow for a breaking of the SU(4) symmetry, the situation 
reverses, with the $U_{\rm MIT}$ being lower at half filling (where it is no longer continuous) than at quarter filling. Moreover, our 
calculations confirm the fact that, when restricting to the case with no magnetic or orbital order, the effect of the Hund coupling $J$ 
at $n=1$ is to shift upward the MIT, as previously suggested by DMFT and slave-particle approaches~\cite{demedici2011a,demedici2011b}. 
In fact, the insulator with one electron per site does not have any substantial advantage from the presence of the Hund coupling, while 
the metallic phase, where the number of double occupancies is higher than in the insulator, gains potential energy when two electrons with 
the same spin are on the same site (and different orbitals). Finally, we would like to mention that, given the very gradual modification 
of the Jastrow factor (and correspondingly the density-density correlations), it is difficult to give a precise determination of 
$U_{\rm MIT}/t$ when considering fully-symmetric wave functions (also for the case with $n=2$ and $J=0$, see above); locating $U_{\rm MIT}/t$ 
with high precision is however beyond the scope of this work.

\subsection{The metal-insulator transitions with magnetic/orbital orders}

\begin{figure}
\includegraphics[width=1.0\columnwidth]{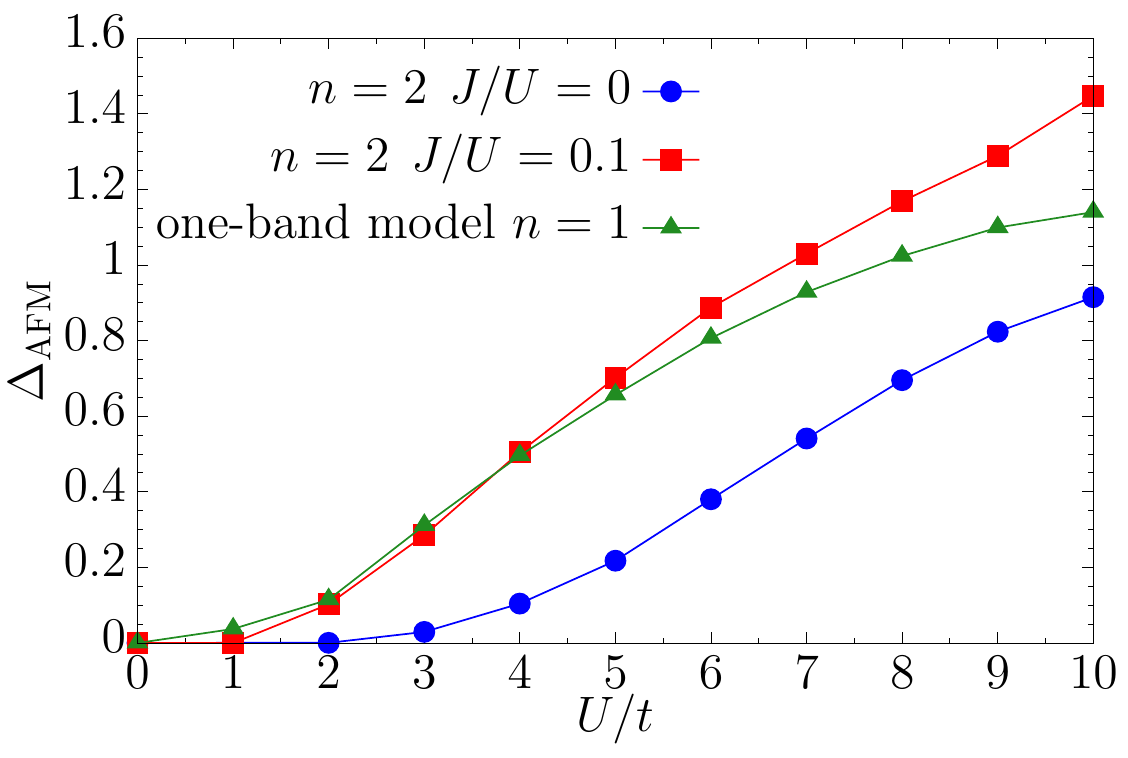}
\caption{\label{fig:deltaAF}
Antiferromagnetic parameter $\Delta_{\rm AFM}$ of Eq.~(\ref{eq:hamMAG}) for $n=2$, as a function of $U/t$. The cases with $J=0$ (full 
circles) and $J/U=0.1$ (full squares) are reported for the two-band Hamiltonian, as well as the results for the single-band Hubbard model 
(full triangles).}
\end{figure}

The above picture for the metal-insulator transitions at $n=1$ and $2$ drastically changes when magnetic and/or orbital order is allowed 
within the non-interacting wave function, i.e., when also the last two terms of Eq.~(\ref{eq:hamaux}) are considered. At half filling, a 
finite (staggered) magnetic order can be clearly stabilized for $J \ge 0$ (while no orbital order is detected). Notice that, in the case 
with $J=0$, magnetic and orbital orders are related by SU(4) symmetry and, therefore, also an orbital order can be found. The optimized 
antiferromagnetic parameter $\Delta_{\rm AFM}$ of Eq.~(\ref{eq:hamMAG}) is reported in Fig.~\ref{fig:deltaAF}, for $J=0$ and $J/U=0.1$. 
In the former case, $\Delta_{\rm AFM}$ is significantly reduced with respect to the single-band model, which is also reported for comparison. 
In the presence of a finite $\Delta_{\rm AFM}$, the triplet pairing $\Delta_{\perp}$ is vanishing (or very small); however, a variational 
wave function with no magnetic order but a finite triplet pairing can be still stabilized as a local minimum at higher variational energies. 
Our results for $\Delta_{\rm AFM}$ are compatible with a finite magnetic order down to $U=0$, with an exponentially small magnetization 
for $U/t \to 0$. Given the smallness of the energy gain due to $\Delta_{\rm AFM}$ in the weak-coupling limit (i.e., $U/t \lesssim 2$), we 
are not able to exclude the possibility that antiferromagnetism sets in at a (small) finite value of $U/t$ and not exactly at $U=0$. 
Nevertheless, our variational calculations clearly support the existence of antiferromagnetism at half filling for intermediate values of 
$U/t$. Moreover, since the SU(4) Heisenberg model with two (fermionic) particles per site is expected to be ordered~\cite{kim2017} and 
since a finite Hund coupling cooperates with the super-exchange mechanism to favor staggered magnetism, we foresee that magnetic order should 
survive for any value of $U/t$ up to $U/t \to \infty$.

%%%%%%%%%%%%%%%%%%%%%%%%%%%%%%%%%%%%%%%%%%%%%

\begin{figure}
\includegraphics[width=1.0\columnwidth]{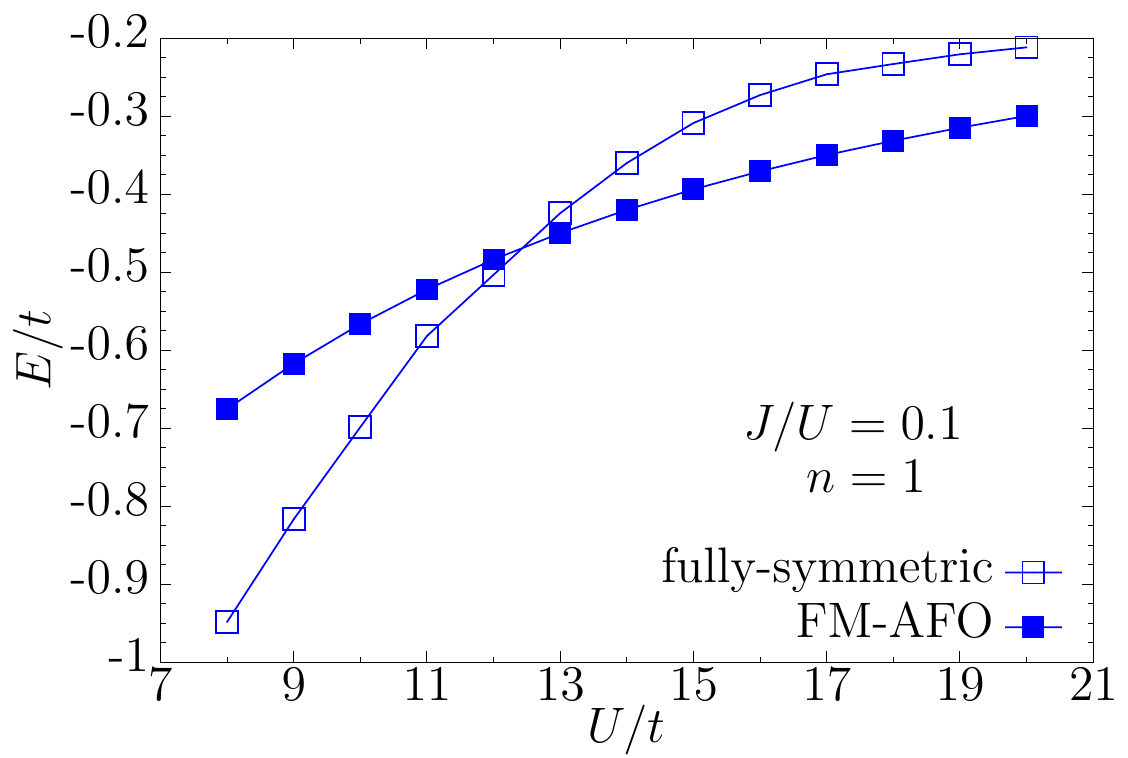}
\caption{\label{fig:orbord}
Variational energies (per site) for the fully-symmetric wave function (empty squares) and the one that contains ferromagnetic (FM) 
and antiferro-orbital (AFO) orders (full squares), for $n=1$ and $J/U=0.1$.}
\end{figure}
%%%%%%%%%%%%%%%%%%%%%%%%%%%%%%%%%%%%%%%%%%%%%

For $n=1$, no evidence for antiferromagnetic order is obtained, at least for $U/t \lesssim 25$. Instead, in the presence of a finite Hund 
coupling, a considerable energy gain is found in the strong-coupling regime by allowing both ferromagnetic and antiferro-orbital order, 
since virtual-hopping processes favor configurations in which two electrons on neighbor sites have parallel spins and reside on different 
orbitals~\cite{kubo2009,peters2010}. Indeed, for sufficiently large electron-electron repulsion, the best variational state is insulating with 
saturated magnetization $m=(n_{\uparrow}-n_{\downarrow})/(n_{\uparrow}+n_{\downarrow})=1$ (where $n_{\sigma}=\sum_{i,\alpha} n_{i,\alpha,\sigma}$) 
and a finite $\Delta_{\textrm{AFO}}$ in Eq.~(\ref{eq:hamORB}). By contrast, for small values of $U/t$, a fully-symmetric metal with $m=0$ and 
no orbital order is found. No intermediate values of $m$ can be stabilized with orbital order. The results for $J/U=0.1$ are reported in 
Fig.~\ref{fig:orbord}, where a first-order phase transition between a metallic state with $m=0$ and no orbital order and an insulator 
with $m=1$ appears at $U/t=12.5 \pm 0.5$.

\subsection{Superconductivity and magnetism}\label{sec:superc}

In the single-band Hubbard model, several calculations have suggested that (singlet) $d$-wave superconductivity emerges upon doping the Mott 
insulating state at half filling~\cite{halboth2000,maier2005,eichenberger2007,gull2013,yokoyama2013,kaczmarczyk2013,deng2015,tocchio2016b}.
Within the resonating valence-bond picture~\cite{anderson1987a,anderson1987b,baskaran1988}, this result can be explained by the existence
of ``preformed'' electron pairs in the Mott insulator, where conduction is impeded by the strong electron-electron repulsion; then, phase 
coherence of mobile pairs emerges upon hole doping. In our variational picture, a necessary condition for having finite superconducting 
correlations is the presence of a finite pairing amplitude in the auxiliary Hamiltonian of Eq.~\ref{eq:hamBCS}. Indeed, in the single-band
model, a finite BCS pairing with $d$-wave symmetry can be stabilized for moderate and large values of $U/t$~\cite{tocchio2016b}. This picture 
becomes less robust in the multiband Hubbard model with degenerate electronic bands. For very small values of the Hund coupling (including 
$J=0$), a finite pairing amplitude $\Delta_d$ with $d_{x^2-y^2}$ symmetry can be stabilized at half filling, see Fig.~\ref{fig:mitn2J0}; 
however, $\Delta_d$ drops to zero for very small dopings, i.e., around $n \approx 1.95$. Singlet pairing is not present at finite doping 
also when different symmetries of the gap function are taken into account; in this respect, we have considered also an extended $s$-wave pairing 
with nearest- and next-nearest-neighbor coupling. In addition, for $J/U \gtrsim 0.1$, no intra-orbital pairing can be stabilized in the wave 
function, even at half filling. We would like to mention that one way to recover a finite singlet pairing at reasonably large dopings is to 
break the symmetry between the inter- and the intra-orbital Coulomb repulsion, e.g., considering $J=J^\prime=0$ but still $U \gg U^\prime$. 
In this case, orbital fluctuations are reduced (since configurations with two electrons on different orbitals are favored over the ones with 
a doubly-occupied orbital) and the resulting physical behavior can be assimilated to the one of two (weakly-coupled) single-band Hubbard models 
(one for each orbital). Therefore, a finite $d$-wave pairing can be stabilized at finite dopings. We also mention that, in the opposite limit 
with $U \ll U^\prime$, an on-site $s$-wave pairing is present close to half filling, since doubly-occupied orbitals are favored over 
singly-occupied ones. Remarkably, these two kinds of pairings compete with each other and no singlet pairing can be stabilized away from half 
filling in the isotropic case with $U=U^\prime$.

\begin{figure}
\includegraphics[width=1.0\columnwidth]{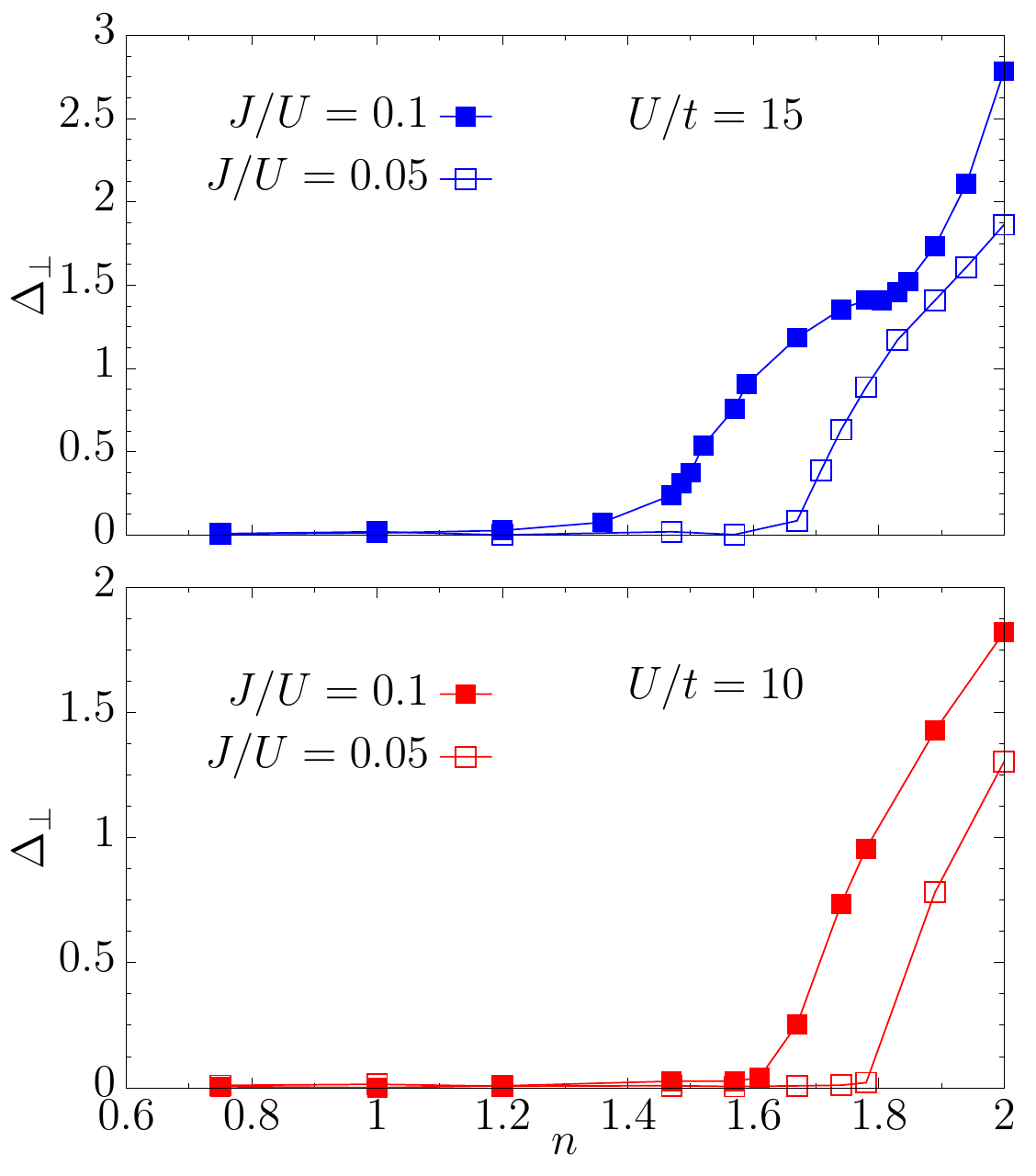}
\caption{\label{fig:SC1}
Triplet pairing $\Delta_{\perp}$ in the auxiliary Hamiltonian of Eq.~(\ref{eq:hamaux}) when no magnetic or orbital order is considered. 
Results are reported for $U/t=15$ (upper panel) and $U/t=10$ (lower panel) for two values of the Hund coupling $J/U=0.05$ (empty squares) 
and $J/U=0.1$ (full squares).}
\end{figure}

When no magnetic and orbital order are allowed in the variational wave function, a sizable interband triplet pairing $\Delta_{\perp}$ is present 
in the vicinity of $n=2$ for $J>0$ and sufficiently large Coulomb repulsion $U$, see Fig.~\ref{fig:SC1}. This is a consequence of the fact that,
on each site, $S=1$ states are favored when $J>0$; a similar feature, with the developments of large local moments, has been also suggested by a 
recent DMFT study of the three-band model~\cite{hoshino2015}. However, in contrast to the latter work, which found that an Ising anisotropy in 
the Hund coupling is important to stabilize triplet superconductivity, we have evidence that a finite triplet pairing is present also in the 
isotropic case, which is modeled by the Hamiltonian of Eq.~(\ref{eq:hamint}). It must be emphasized that, away from half filling, the presence 
of a finite electron pairing in the uncorrelated wave function implies a true long-range order, since the Jastrow pseudopotential has 
$v^{\rm intra}(\bm{q}) \approx v^{\rm inter}(\bm{q}) \propto 1/|\bm{q}|$. As expected, the strength of triplet superconductivity is proportional 
to the Hund coupling, thus implying that the doping region in which $\Delta_{\perp} \ne 0$ enlarges with increasing $J$, see Fig.~\ref{fig:SC1}.

%%%%%%%%%%%%%%%%%%%%%%%%%%%%%%%%%%%%%%%%%%%%%

\begin{figure}
\includegraphics[width=1.0\columnwidth]{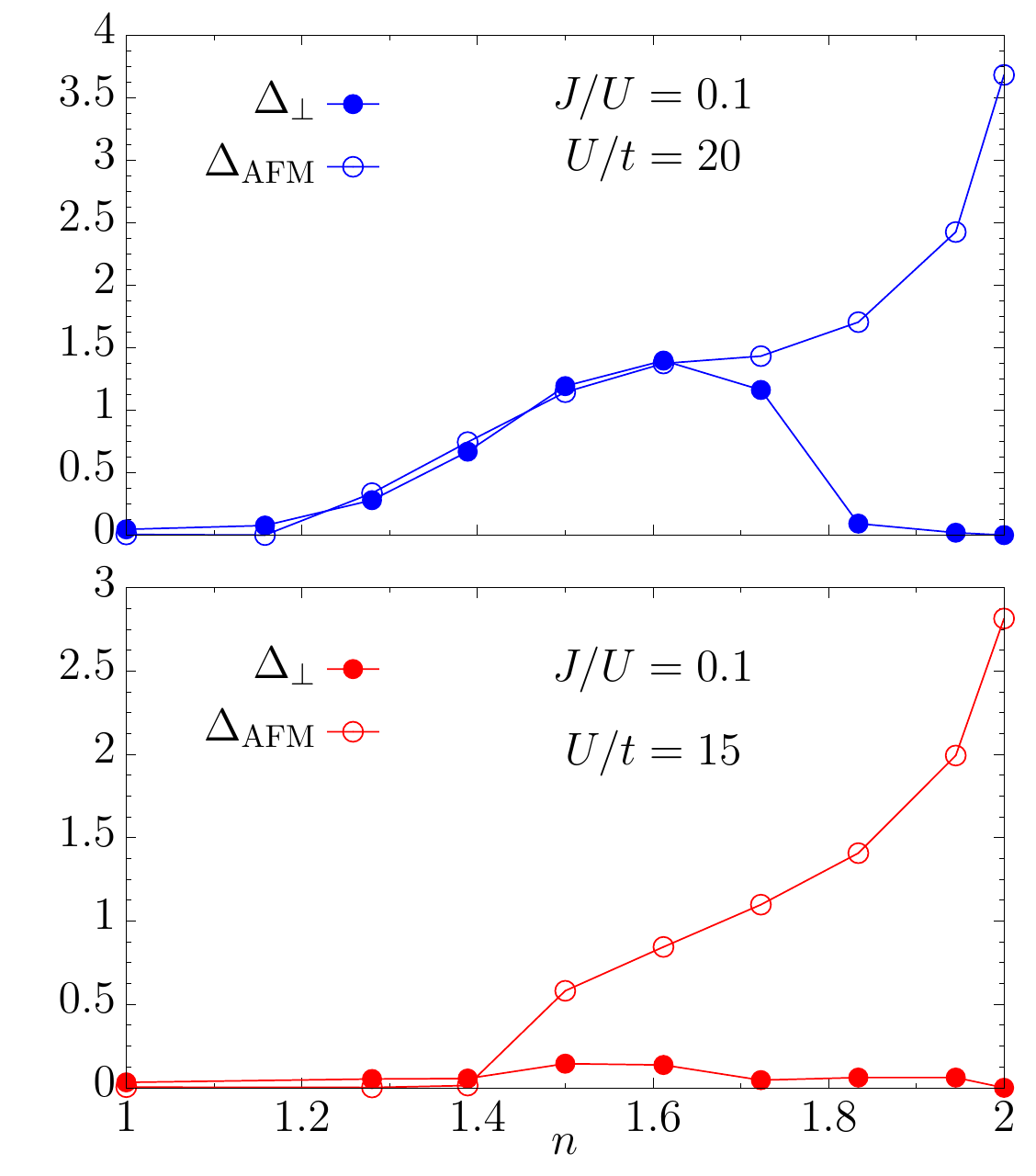}
\caption{\label{fig:SC2}
Triplet pairing $\Delta_{\perp}$ (full circles) and antiferromagnetic order parameter $\Delta_{\rm AFM}$ (empty circles) in the auxiliary 
Hamiltonian of Eq.~(\ref{eq:hamaux}). Results are reported for $U/t=20$ (upper panel) and $U/t=15$ (lower panel), for $J/U=0.1$.}
\end{figure}
%%%%%%%%%%%%%%%%%%%%%%%%%%%%%%%%%%%%%%%%%%%%%

\begin{figure}
\includegraphics[width=1.0\columnwidth]{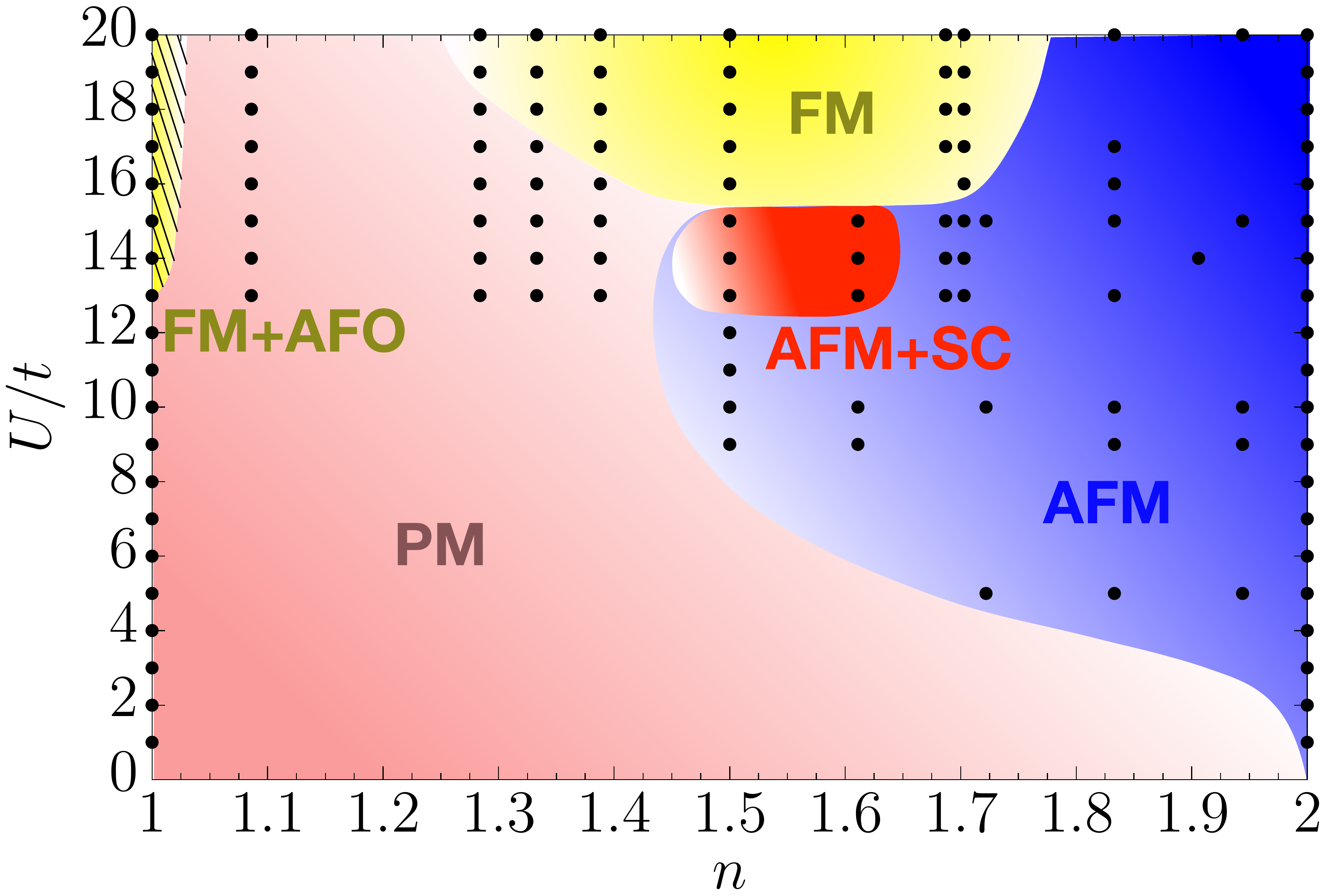}
\caption{\label{fig:pd}
Schematic phase diagram of the two-band Hubbard model in the $(n,U/t)$ plane, for $J/U=0.1$. The yellow region denotes ferromagnetism (FM), which is
expected to possess also antiferro-orbital order (AFO) in a tiny region close to $n=1$ (shaded region). The blue region has an antiferromagnetic ground 
state (AFM), while the red one shows a coexistence of antiferromagnetism and superconductivity with triplet pairing (AFM+SC). Finally, the pink region is a 
paramagnetic metal (PM). The concomitant presence of ferromagnetism and superconductivity is not investigated. Data, shown as black points, are obtained 
on clusters with $12 \times 12$, $16 \times 16$, and $18 \times 18$ sites.}
\end{figure}

When also magnetism is included in the variational wave function, superconductivity is largely suppressed. First of all, antiferromagnetic 
correlations are strong for electron densities close to half filling. Here, we can consider wave functions that contain both electron pairing
and antiferromagnetism and optimize $\Delta_{\perp}$ and $\Delta_{\rm AFM}$ together. The results are shown in Fig.~\ref{fig:SC2} for $J/U=0.1$. 
When $\Delta_{\rm AFM}$ is present, triplet pairing is strongly reduced close to half filling, leading to an antiferromagnetic metal with no 
pairing correlations. For $U/t=15$, a tiny triplet superconductivity emerges around $n=1.5$, where antiferromagnetism is still present, thus 
leading to a coexistence between these two order parameters. The pairing amplitude becomes much stronger when increasing the value of the Coulomb 
interaction, e.g., for $U/t=20$, where $\Delta_{\perp}$ displays a dome-like feature with a broad maximum at $n \approx 1.6$. However, in the 
presence of a finite Hund coupling also ferromagnetism becomes competitive in energy, especially when $U/t$ is large. A direct comparison between 
the superconducting state (with or without antiferromagnetic order) and the ferromagnetic one (with or without orbital order) allows us to draw 
the phase diagram of Fig.~\ref{fig:pd} for $J/U=0.1$. In this case, the best variational state has antiferro-orbital order for $n=1$, while a 
uniform ferromagnet exists in a wide region at finite electron densities and large $U/t$. Instead, close to $n=1$, a paramagnetic metal intrudes 
between these two ferromagnetic states. Our results are in qualitative agreement with previous variational~\cite{kubo2009} and DMFT~\cite{peters2010} 
calculations, which found the existence of uniform ferromagnetism at large values of the Coulomb repulsion for $1<n<2$. Orbital order should survive 
in a tiny region close to quarter filling; however, even on the largest cluster that we considered (i.e., $18 \times 18$) at $n \approx 1.1$ (which 
is the closest available density to quarter filling that allows a direct comparison between ferromagnetic and paramagnetic states) the ferromagnetic 
wave function has a slightly higher energy than the paramagnetic one. For $J/U=0.1$, phase separation is expected to take place close to the 
transition between the paramagnetic and the ferromagnetic metals, because of the first order nature of the transition. We remark that this is 
conceptually different from the scenario proposed in Ref.~\cite{demedici2017b}, where the paramagnetic metal acquires a diverging susceptibility
when approaching half filling. For larger values of $J/U$, the ferromagnetic state can be stabilized also close to quarter filling, thus eliminating 
phase separation (not shown). The possibility to have triplet superconductivity inside the ferromagnetic region could be investigated by using 
Pfaffian wave functions~\cite{spanu2008}, in which pairing is considered for electrons with parallel spins. This kind of approach goes well beyond 
the scope of the present work.

%%%%%%%%%%%%%%%%%%%%%%%%%%%%%%%%%%%%%%%%%%%%%
%%%%%%%%%%%%%%%%%%%%%%%%%%%%%%%%%%%%%%%%%%%%%

\section{Conclusions}\label{sec:conclusions}

We have considered the two-band Hubbard model with degenerate electronic bands by using variational wave functions and Monte Carlo techniques. 
At integer fillings with $n=1$ and $n=2$, we have first investigated the metal-insulator transitions when both magnetic and orbital order are not 
considered. In this regime, our results for the location of the MIT, as a function of the Hund coupling $J$, are qualitatively in agreement with 
previous DMFT and slave-particle approaches~\cite{demedici2011a,demedici2011b}. At half filling for $J>0$, the transition is first (second) order 
for small (large) values of the Hund coupling, with a sizable triplet pairing within the Mott insulator (still, no superconducting long-range 
order is established at half filling, because of the strongly repulsive Jastrow factor). At quarter filling, the transition is second order with 
no finite pairing neither in the metallic nor in the insulating phase. 

We have then included the possibility to stabilize magnetic and/or orbital order. At half filling, a clear evidence for antiferromagnetic order 
has been obtained for $J \ge 0$. In particular, the qualitative behavior of the magnetic parameter resembles the one of the single-band Hubbard 
model, where antiferromagnetic order sets in at $U=0$; therefore, our results suggest that the ground state for $n=2$ is antiferromagnetically 
ordered for any positive value of the Coulomb interaction $U$. Triplet pairing is not present when a finite antiferromagnetic parameter is stabilized.
At quarter filling, no sign of antiferromagnetic order is detected (up to $U/t=25$); instead for $J>0$, the ground state shows a first-order phase 
transition from a metallic state for small values of the electron-electron interaction to an insulator with staggered orbital order and ferromagnetic 
correlations in the strong-coupling regime. 

At intermediate electron dopings with $1<n<2$, when both magnetic and orbital order are not included, a sizable triplet pairing is present for finite 
values of the Hund coupling and sufficiently large electron-electron interactions, i.e., when the Mott insulator at $n=2$ is doped. A similar trend
has been recently found by DMFT calculations on the three-band Hubbard model~\cite{hoshino2015}. However, in our case, the Ising anisotropy in the 
Hund coupling is not necessary to obtain triplet pairing. We report that, at odds with the single-band Hubbard model, no sizable singlet pairing 
is instead present away from $n=2$. When magnetic order is also considered within the variational wave function, triplet superconductivity is 
strongly suppressed by antiferromagnetic order close to $n=2$; furthermore, the region where superconductivity can be stabilized is also reduced by 
the presence of ferromagnetism, which is competitive in a wide range of densities for large Coulomb repulsions. The possibility to have a coexistence 
of triplet pairing and ferromagnetism could be considered by extending our variational approach to Pfaffian states, which is however quite expensive 
for multiband models and goes beyond the scope of this work.

\acknowledgments

We thank M. Fabrizio and L. de' Medici for useful discussions.

\end{document}